\documentclass[twocolumn]{emulateapj}

\newcommand{\msun}{M_{\odot}}

\newcommand{\avsbeff}{\langle \mu \rangle_{e}}

\accepted{October 13, 2017}

\shorttitle{Distance and Properties of NGC 4993}
\shortauthors{Im et al.}

\begin{document}

\title{Distance and properties of NGC 4993 as the host galaxy of a gravitational wave source, GW170817}

\author{Myungshin Im\altaffilmark{1,7}, Yongmin Yoon\altaffilmark{1}, Seong-Kook Lee\altaffilmark{1},
Hyung Mok Lee\altaffilmark{2}, Joonho Kim\altaffilmark{1}, Chung-Uk Lee\altaffilmark{3}, 
Seung-Lee Kim\altaffilmark{3}, Eleonora Troja\altaffilmark{4,5}, Changsu Choi\altaffilmark{1},
Gu Lim\altaffilmark{1}, Jongwan Ko\altaffilmark{3}, Hyunjin Shim\altaffilmark{6}}

\affil{\altaffilmark{1}Center of the Exploration of the Origin of the Universe, Astronomy Program, 
Dept. of Physics \& Astronomy, Seoul National University, 
1 Gwanak-rho, Gwanak-gu,
Seoul, 08826, Republic of Korea}
\affil{\altaffilmark{2}Astronomy Program, 
Dept. of Physics \& Astronomy, Seoul National University, 
1 Gwanak-rho, Gwanak-gu,
Seoul, 08826, Republic of Korea}
\affiliation{\altaffilmark{3}Korea Astronomy and Space Science Institute, 776 Daedeokdae-ro, Yuseong-gu, Daejeon 34055, Korea}
\affiliation{\altaffilmark{4}Department of Astronomy, University of Maryland, College Park, Maryland 20742-4111, USA}
\affiliation{\altaffilmark{5}NASA Goddard Space Flight Center, 8800 Greenbelt Rd, Greenbelt, Maryland 20771, USA}
\affiliation{\altaffilmark{6}Department of Earth Science Education, Kyungpook National University, 80 Daehak-ro, Buk-gu, Daegu 41566, Republic of Korea}
\email{\altaffilmark{7}mim@astro.snu.ac.kr}



\begin{abstract}

Recently, the optical counterpart of a gravitational wave source GW170817 has been identified in NGC 4993 galaxy. 
Together with evidence from observations in electromagnetic waves, the event has been suggested as a result of a merger of two neutron stars.  
We analyze the multi-wavelength data to characterize the host galaxy property and its distance to examine if the properties of NGC 4993  are consistent with this picture. Our analysis shows that NGC 4993 is a bulge-dominated galaxy with $r_{\rm eff} \sim 2-3$ kpc and the S\'ersic index of $n = 3-4$ for the bulge component. The spectral energy distribution from 0.15 to 24 $\mu$m indicates that this galaxy has no significant ongoing star formation, the mean stellar mass of 
$(0.3 - 1.2) \times 10^{11} \, \msun$,
the mean stellar age greater than $\sim$3 Gyr, and the metallicity of about 20\% to 100\% of solar abundance.
 Optical images reveal dust lanes and extended features that suggest a past merging activity. Overall, NGC 4993 has characteristics of normal, but slightly disturbed elliptical galaxies. Furthermore, we derive the distance to NGC 4993 with the fundamental plane relation using 17 parameter sets of 7 different filters 
  and the central stellar velocity dispersion from literature, 
 finding an angular diameter distance of $37.7 \pm 8.7$ Mpc. NGC 4993 is similar to 
 some host galaxies of short gamma-ray bursts but much different from those of long gamma-ray bursts, supporting the picture of GW170817 as a result of a merger of two NSs.

\end{abstract}


\section{Introduction} \label{sec:intro}

 Since 2015, the Advanced Laser Interferometer Gravitational-Wave Observatory (LIGO) and the Advanced Virgo have succeeded in detecting a number of GW signals coming from distant black hole (BH) merger events \citep{Abbott16,Abbott17a}.  However, the detected GW events so far have resulted from mergers of two stellar mass BHs that are unlikely to produce detectable optical signals, 
and no optical sources have been identified that correspond to the GW sources. A merger of two neutron stars (NSs) or a NS and a BH has been suggested as a GW source that produces optical signal. Indeed, NS mergers are thought to be responsible for short gamma-ray bursts (GRB) when viewed on-axis, and models have predicted an optical signal, known as kilonova, to be produced in such an event when viewed off-axis \citep[e.g., see a review by][and references therein]{Metzger17}.
 
  GW170817 is a GW source whose signal was detected by the Advanced LIGO and the Advanced Virgo on 2017 August 17, 12:41:04 UT \citep{Abbott17b}.  A weak and short ($\sim$2 sec) gamma-ray signal was caught by the Fermi Gamma-ray Space Telescope and INTEGRAL two seconds after the GW detection \citep{Goldstein17, Savchenko17}. The GW signal showed that this could be a merger of two NSs at a
 distance of $40^{+8}_{-14}$ Mpc \citep{Abbott17b}. 
  About 11 hours after the LIGO detection, an optical counterpart of the GW source was reported \citep{Coutler17}, in a nearby galaxy NGC 4993. The source has been found to have characteristics of off-axis short GRB and a kilonova \citep[e.g.,][]{Troja17}.
  
  If GW170817 is caused by a binary NS (BNS) merger, the host galaxy is expected to have properties similar to the hosts of short GRBs.  
 So far, host galaxies of short GRBs are known to have ages in the range of several tens Myr to $\sim$5 Gyr, stellar masses of $10^{8.5 - 11.8} \, \msun$ (a median value of $10^{10}\, \msun$), zero to $\sim$10 $\msun$ yr$^{-1}$ star formation rate, a median metallicity of solar abundance and morphology of both spirals and ellipticals \citep{Berger14,Leibler10,Troja16}. This is very different from hosts of long GRBs that are younger, actively star-forming, less metal-rich, and less massive \citep[e.g.,][]{Michalowski12}.  

 The host galaxy, NGC 4993, has been known as an early-type galaxy in the ESO 508 cluster.
 The distance to ESO 508  has been measured to be 41.1 Mpc using the Cepheid-calibrated Tully-Fisher relation on the cluster spiral galaxis \citep{Sakai00}, but this is by no means direct measurement of distance to NGC 4993.
  Furthermore, while analysis of the GW signal provides an independent measurement of the luminosity distance, the distance based on GW can suffer from a non-negligible amount of uncertainties mostly due to the uncertainties in the inclination angle of the orbital plane with respect to the line of sight \citep{Abbott16}. If the distance is accurately measured by other means, the inclination angle of the orbital plane can be well constrained. In that sense, accurate measurements of the distances to the host galaxies are very important. 

   About 8 hours after the first identification of the optical counterpart, we started follow-up observation of GW170817 (Im et al. 2017, in preparation; Troja et al. 2017). The accumulated dataset allows us to construct deep images that can unveil faint, extended features and to accurately determine physical parameters of NGC 4993. Using these images, as well as other multi-wavelength data, we study properties of NGC 4993 to examine if this galaxy has the characteristics of short GRB hosts, and derive an independent measure of its distance. 
To obtain the physical sizes, we adopt the angular diameter distance of 37.7 Mpc from this paper. The magnitudes are in AB system.

\section{Data} \label{sec:data}

  
\subsection{KMTNet and LSGT observations}   

  We observed NGC 4993 from 2017 August 18 through September 7 at three locations, the Siding Spring Observatory (SSO) in Australia, the South African Astronomical Observatory (SAAO) in South Africa, and the Cerro-Tololo Inter-American Observatory (CTIO) in Chile, using 1.6m telescopes of the Korean Microlensing Telescope Network (KMTNet; Kim et al. 2016).  Images were taken in $B, V, R$, and $I$ filters, and the data are reduced with the standard KMTNet pipeline. These data are stacked to create deep images with the total integration times of 1140 s, 1260 s, 7760 s, and 9900 s for $B, V, R$, and $I$ filters respectively. The surface brightness (SB) limit reaches to 27.9 mag arcsec$^{-2}$ in $R$-band (1-$\sigma$) when the pixels\footnote{KMTNet pixel scale is $0\farcs396$.} are binned by $9 \times 9$. We calibrated the photometry using the AAVSO Photometric All-Sky Survey (APASS) stars within 30$\arcmin$ from NGC 4993 at a magnitude range of 14 to 17.  When stacking images, we used background-subtracted images for which the flux scales are normalized. The background subtraction was carefully done by choosing a large background estimation kernel (10$\arcmin$ or larger) so that the background does not get over-subtracted during this process.
  
  We also observed NGC 4993 using the SNUCAM-II \citep{Choi17} on the 0.43m Lee Sang Gak Telescope \citep[LSGT;][]{Im15} and 0.5m class telescopes of the iTelescope.Net at SSO from 2017 August 18 through September 11. For the observation, we used the $u, g, r, i, z$, a set of medium-band filters, and $B,V,R,I$ filters. The $u$-band image  has one hour on-source integration, and it is used to derive the $u$-band flux for SED fitting. The photometry calibration was done by using the APASS and the Two Micron All Sky Survey (2MASS) data of stars near NGC 4993 as described in \citet{Choi17}.


\begin{figure*}[!hpt]  
\includegraphics[scale=1.25]{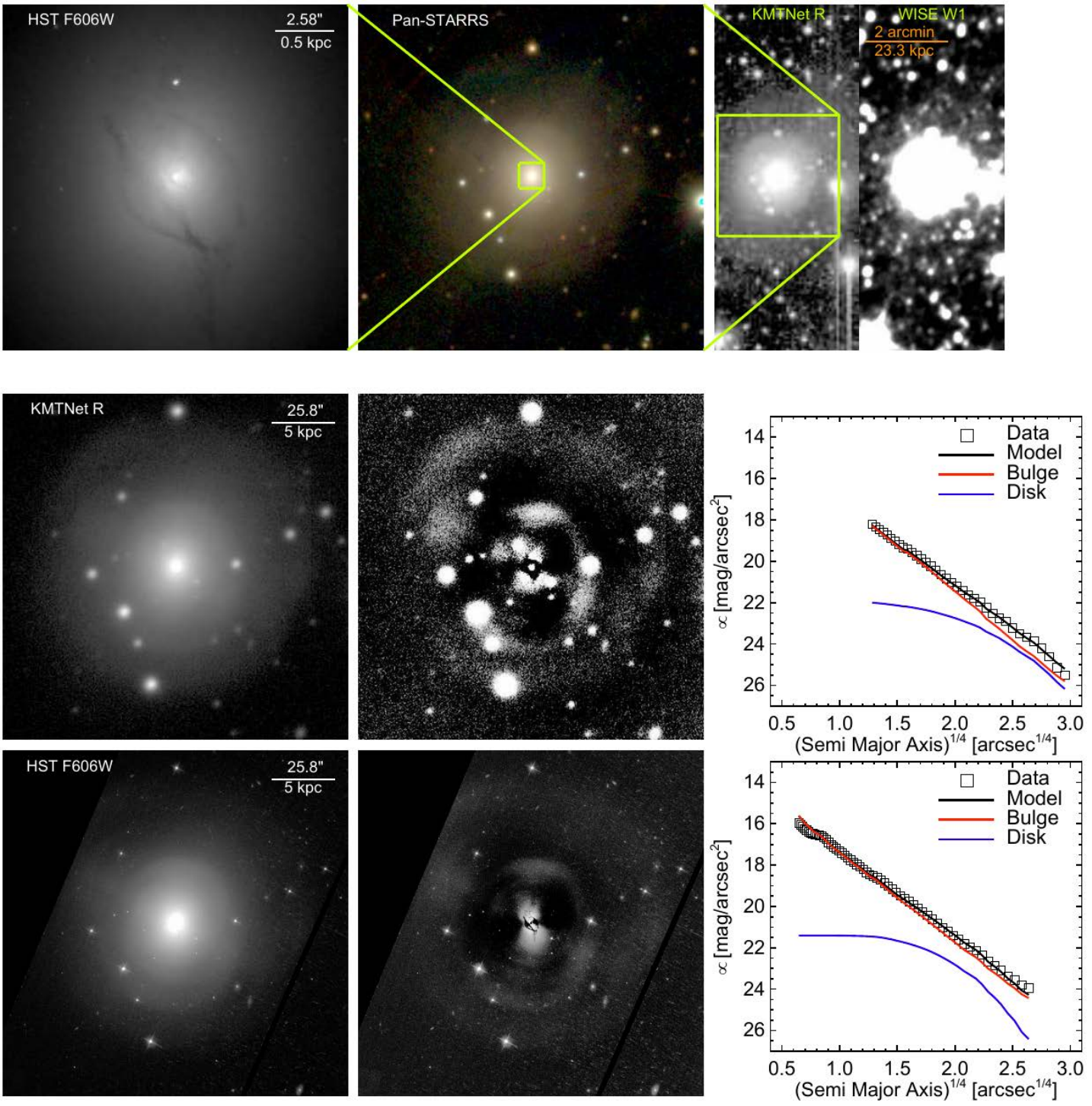}
\caption{Images of NGC 4993 and the result of the SB fitting. Top panels show NGC 4993 in the innermost region (HST $F606W$), the region over several $\theta_{\rm eff}$ radii (the Pan-STARRS color composite), and the outer region that extends out to 8 $\theta_{\rm eff}$ radii ($9\times9$ binned KMTNet $R$-band and WISE $W1$ images). The $F606W$ image reveals dust lanes in the inner part of NGC 4993, while the Pan-STARRS and the KMTNet images show extended features in the outer part. In the second and the third rows, we show the SB fitting results for the $R$, and the $F606W$ images. From left to right, we show the original image, the residual image after the model image subtraction, and the 1-dimensional SB profile of the best-fit model in comparison to the data.   The 1-dimensional SB profile shows that NGC 4993 has a SB with a S\'ersic index around $n=4$, a typical value for elliptical galaxies.\label{fig:f1}}
\end{figure*}

\subsection{Archive data}
 
  GALEX All-sky Imaging Survey observed NGC 4993 in far- and near-ultraviolet (FUV and NUV), and the stacked, calibrated images are taken from the GALEX data archive.   Coadded $g, r, i, z$, and $y$ filter images are obtained from the Panoramic Survey Telescope and Rapid Response System (Pan-STARRS) Data Release 1 \citep{Chambers16}.  We also used the publicly available F606W image taken by the HST Advanced Camera for Surveys (ACS).   In infrared, the $J, H,$ and $K_{s}$ filter images are taken from 2MASS,  and we also utilized the WISE $W1, W2, W3$, and $W4$ images.



\section{Morphology and Structural Parameters}\label{sec:morphology}

  Fig. \ref{fig:f1} shows the stacked KMTNet $R$, the HST $F606W$, Pan-STARRS $g, r, i$ color composite, and the WISE $W1$ band images. In the inner region of a scale of about 1 kpc, we see dust lanes and a nuclear part that appears to have a dust-obscuring central disk. At an outer region of 10 kpc scale, we find several layers of extended features that are commonly regarded as remnant feature of past merger activities \citep{Kim13}. Even further out, we find that the outermost part of the galaxy extends out to 2$\arcmin$ in radius 
  (22 kpc or 8 times of the angular effective radius $\theta_{\rm eff}$\footnote{This is a fitted parameter from the SB-fitting, converted to a circularized value based on the model fit.}), which is visible in both the 9$\times$9 binned $R$-band image and the $W1$ image. This shows that a special care must be taken when stacking images and fitting a SB profile, so that the extended low SB feature is not washed away. 

  We performed the 2-dimensional SB fitting using the Galfit software (Peng et al. 2010), on the $B,V,R,I, F606W, J, H$, and $K_{s}$ images. The point spread functions are constructed using stars in the vicinity of NGC 4993 in each band.  Several models have been used, such as (i) a single S\'ersic component model; (ii)  a two-component model with a S\'ersic profile and an exponential profile; and (iii) a double S\'ersic-component model. We also performed a growth curve analysis, to derive $\theta_{\rm eff}$ of NGC 4993 independently. 
    
  The result of the 2-dimensional SB fitting is summarized below and in Table \ref{tab:sbfit}. The three models can fit the observed SB profile almost equally well, with the two component models at a slightly better $\chi^{2}_{\rm red}$
 (e.g., $\chi^{2}_{\rm red}=0.985$ (single) vs 0.962 (two-component)).  However, the single S\'ersic component model returns $\theta_{\rm eff}$ values about 15\% larger than the two component models. The growth curve analysis of the KMTNet data agrees with the two component model results, and we attribute the larger $\theta_{\rm eff}$ from the single S\'ersic models to an inherent nature of a single S\'ersic profile with a large $n$ that tends to distribute its light to regions at $r \gg \theta_{\rm eff}$. 
 Therefore, we show the result from the S\'ersic + exponential profile model only. In the single S\'ersic model, the S\'ersic index $n$ is found to be 4 to 5, a value common for elliptical galaxies. With a S\'ersic bulge + exponential disk model, we find $B/T \gtrsim 0.7$ with $n =$ 3.0 to 4.4 for the S\'ersic bulge.  The derived $\theta_{\rm eff}$ values correspond to physical effective radii of $r_{\rm eff} = 2 - 3$ kpc. 
   
\begin{deluxetable*}{cccccccccccc}[ht!]
\tabletypesize{\scriptsize}
\tablecaption{SB fitting result\label{tab:sbfit}}
\tablecolumns{12}
\tablenum{1}
\tablewidth{0pt}
\tablehead{
\colhead{Filter} &
\colhead{Bulge mag} & \colhead{Bulge $\theta_{\rm eff}$\tablenotemark{a}} & \colhead{$n$} &
\colhead{Bulge $b/a$} & \colhead{Disk mag} & \colhead{Disk $\theta_{\rm eff}$\tablenotemark{a}} & \colhead{Disk $b/a$}  & 
\colhead{Total mag} & \colhead{Total $\theta_{\rm eff}$\tablenotemark{a}}  & \colhead{B/T} & \colhead{$\avsbeff$} \\
\colhead{} & 
\colhead{(AB)} & \colhead{(arcsec)} & \colhead{} & \colhead{} & 
\colhead{(AB)} & \colhead {(arcsec)} &  \colhead{} & \colhead{(AB)} & \colhead{(arcsec)} &
\colhead{}  & \colhead{(mag/arcsec$^2$)} \\
}
\startdata
B        &   13.6  &  9.8  & 3.0 & 0.83 &  14.5     & 32.3 & 0.83   &  13.2    &  15.4 & 0.68  &  21.15 \\
V        &   12.5  & 11.8  & 3.6 & 0.83 &  14.1     & 35.8 & 0.69   &  12.3    &  15.6 & 0.81  &  20.28 \\
R        &   12.2  & 11.0  & 3.9 & 0.84 &  13.6     & 30.3 & 0.75   &  11.9    &  14.9 & 0.78  &  19.81 \\
I         &   12.0  &  8.6   & 3.5 & 0.84 &  12.8     & 30.4 & 0.82  &   11.5    &  14.2 & 0.68  &  19.29 \\
F606W & 12.4  &  14.8  & 4.6 & 0.88 &  14.2   & 15.8 & 0.95  &   12.2   &  15.1 & 0.84  &   20.11 \\
J        &   11.3  &  7.9   & 4.3 & 0.79 &  12.3     & 22.1 & 0.84  &   10.9    &  12.0 & 0.72  &  18.32 \\  
H       &   11.1 &   7.2   & 4.0 & 0.79 &  12.2     & 23.1 & 0.66  &   10.7    &  10.8 & 0.75  &  17.90 \\      
K       &    11.0 &   9.2   & 4.4 & 0.82 &  13.7     & 21.0 & 0.14  &  10.9     &  10.3 & 0.92  &  18.03 \\
\enddata
\tablenotetext{a}{Circularized effective radii. To get the major axis value, multiply it by $\sqrt{a/b}$.}
\tablecomments{Apparent magnitudes and SBs are not corrected for the Galactic extinction.}
\end{deluxetable*}
 
\begin{deluxetable*}{ccccccc}[hb!]
\tablecaption{Host Galaxy Property from SED-fitting\label{tab:sed}}
\tablefontsize{\scriptsize}
\tablecolumns{7}
\tablenum{2}
\tablehead{
\colhead{} & \colhead{$t$} & \colhead{$\tau$} & \colhead{$M_{*}$} & 
\colhead{$E(B-V)$} & $SFR_{100Myr}$ & $Z$ \\
\colhead{} & \colhead{(Gyr)} & \colhead{(Gyr)} & \colhead{($M_{\odot}$)} & 
\colhead{} & \colhead{($M_{\odot}$ yr$^{-1}$)} & \colhead{} \\
}
\startdata
Best-fit & 6.5 & 0.5 & 3.80e+10 & 0.000 & 4.1e-3 & 0.02 \\
2nd best-fit & 4.25 & 0.3 & 2.93e+10 & 0.025 & 1.9e-3 & 0.02 \\
10Gyr model & 10.0 & 0.5 & 5.48e+10 & 0.075 & 8.7e-6 & 0.008 \\
Piovan & 12.0 & \nodata & \nodata & 0.025 & 0.0 & 0.008 \\
\enddata
\end{deluxetable*}

\section{Stellar Population}\label{sec:stars}

\begin{figure}[hht!]
\includegraphics[width=8cm]{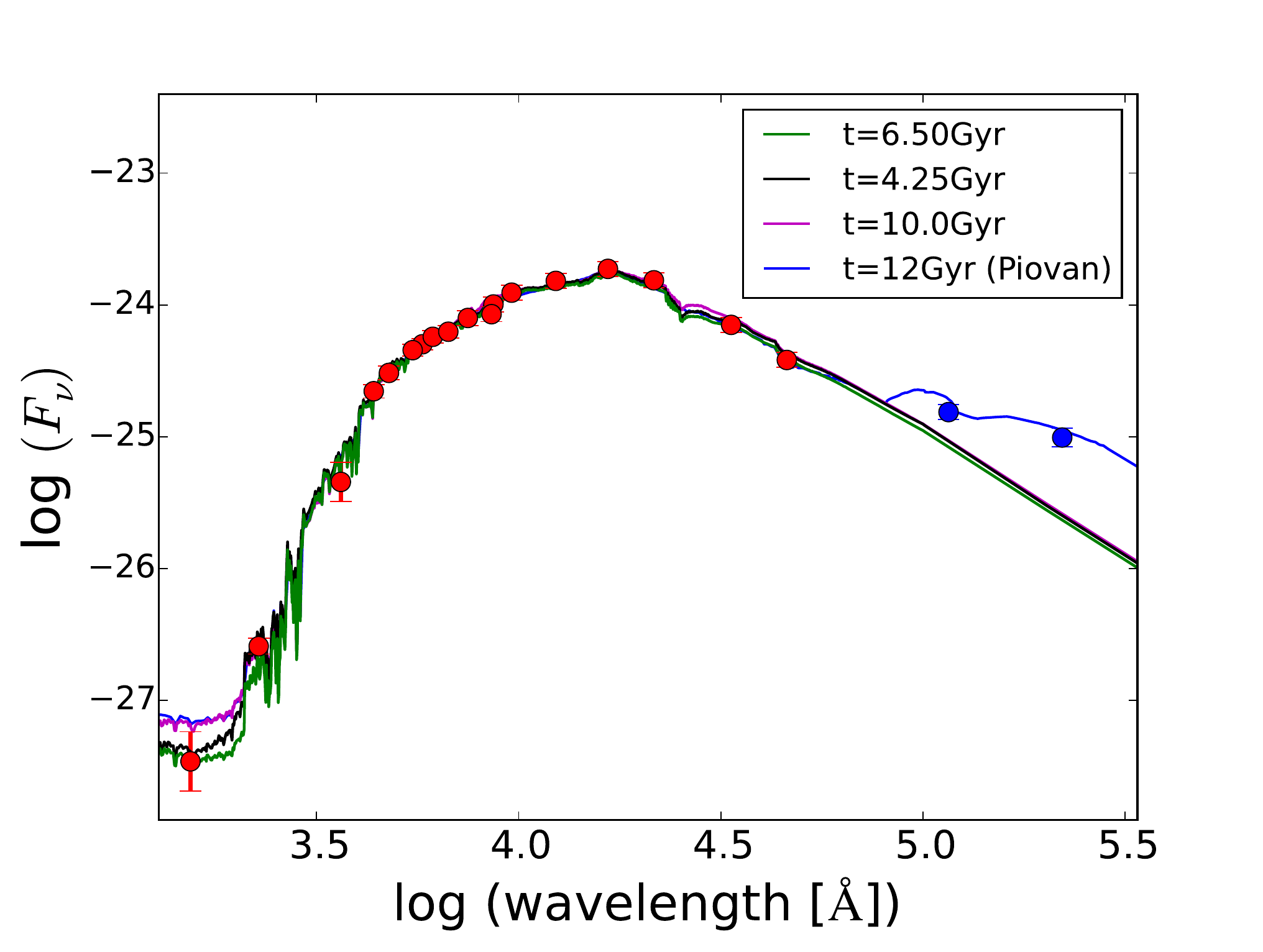}
\includegraphics[width=8cm]{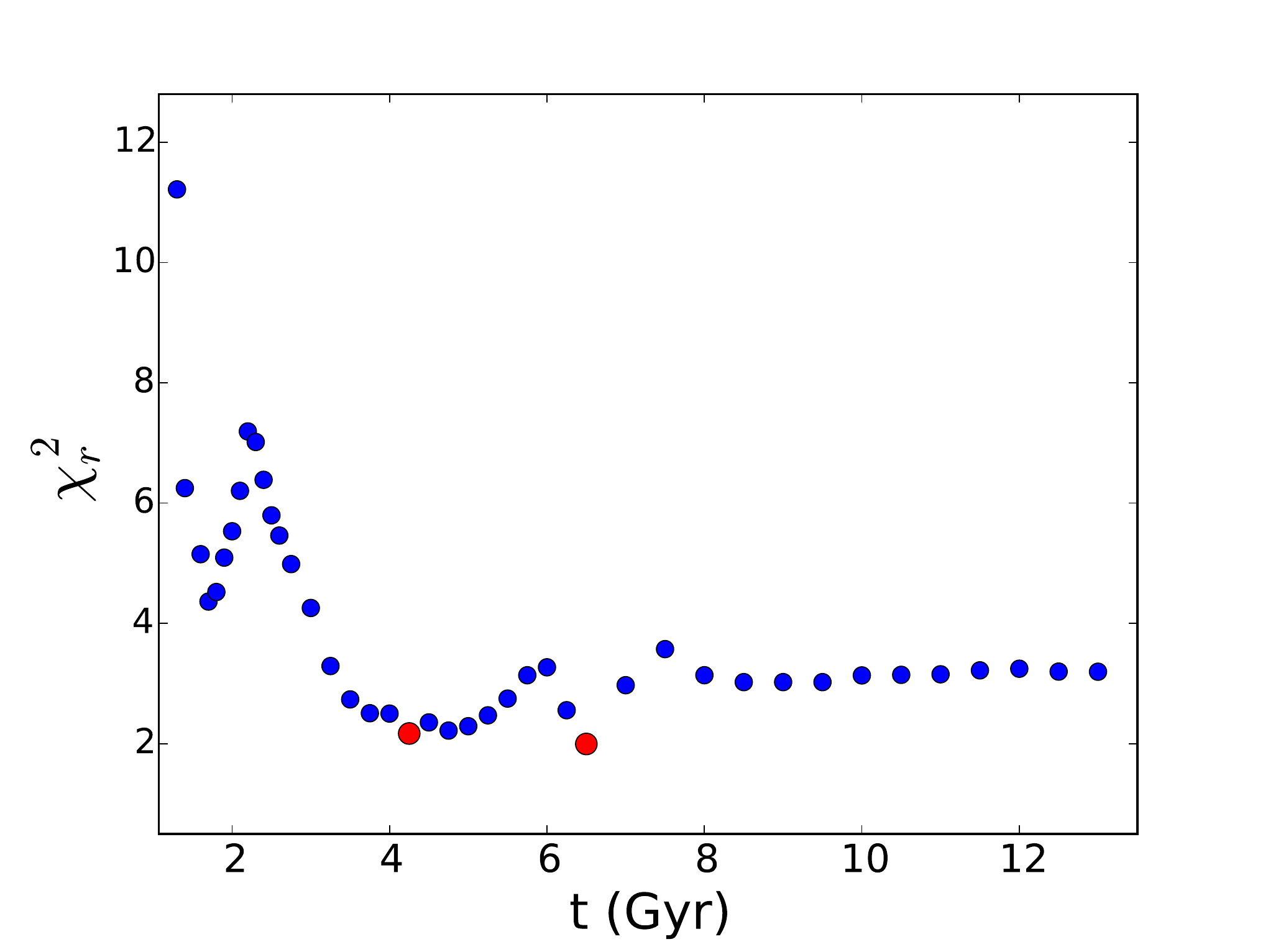}
\caption{(Top) The broad-band SED of NGC 4993 in $FUV, NUV, u,g,r,i,z,y, B, V, R, I, J, H, K, W1, W2, W3$, and $W4$. Overplotted are four SED models that provide a reasonably good match to the data points, among which the best-fit model is the model with $t=6.5$ Gyr.
The Piovan model includes the dust emission from AGB stars; 
(Bottom) The reduced $\chi^{2}$ distribution as a function of the mean stellar age shows two local minima, and a long tail with a reasonably good fit at $t > 3$ Gyr. \label{fig:f2}}
\end{figure}

   We fitted the multi-band photometric data points from FUV to MIR (up to $W2$) using an SED-fitting software of \citet[][and references therein]{Lee15}. This SED model utilizes the stellar population synthesis model of \citet{Bru03} with Padova 1994 stellar evolutionary tracks, the \citet{Cha03} 
initial mass function with a stellar mass range of 0.1 to 100 $\msun$, and a star formation rate in the form of $\frac{t}{\tau^{2}} e^{-t/\tau}$ where $t$ is time since the onset of the star formation, and $\tau$ is the timescale parameter. To fit the data, we used the flux within an aperture of a $30\arcsec$ diameter (5.5 kpc) which was chosen to avoid possible bias due to missing fluxes from the outer, low SB features in shallower images of some filters. The Galactic extinction was corrected by adopting $E(B-V)=0.106$ based on the extinction map of \citet{Schlafly11}, and the extinction curve with $R_{V}=3.1$ of \citet{Fitz99}. The procedure fits five parameters: the age, $\tau$, the metallicity, $E(B-V)$ and the stellar mass.  
  The best-fit stellar mass is scaled up by using the ratio of the total flux to the 30$\arcsec$ diameter aperture flux in $H$-band.  
There is a slight excess at 12 and 24 $\mu$m ($W3$ and $W4$) over the stellar radiation. Therefore we also tried to fit the SED with SSP model templates that include AGB dust emission \citep{Piovan03}, and added the $W3$ and $W4$ data points in the fitting.  
  The fitting results are presented in Fig. \ref{fig:f2} and Table \ref{tab:sed}. Note that a similar result is also presented in \citet{Troja17}, but here we try the fitting with the updated data and include the AGB-dust emission model.

 This SED-fitting results indicate that NGC 4993 has a stellar mass of (3 to 6) $\times \, 10^{10}$ $\msun$,  the metallicity of 20 \% to 100 \% (with the best-fit at 100 \%) and the $\tau$ of 0.3 to 0.5 Gyr. The age is loosely constrainted to be $\gtrsim 3$ Gyr (95 \% confidence). 
 However, the stellar mass is sensitive to the assumed IMF. If we assume the Salpeter IMF instead, the stellar mass becomes about two times larger. Since both the Salpeter and the Chabrier IMFs are plausible \citep{Cappellari12}, we conclude that NGC 4993 has the mean stellar mass in the range of (0.3 to 1.2) $\times \, 10^{11}$ $\msun$.
 The star formation rate is very low, at a value of $\sim 4 \times 10^{-3} \msun$ yr$^{-1}$ or less, which is consistent with no star formation activity. A small amount of internal dust extinction ($E(B-V)=0.025 - 0.075$) is consistent with the presence of the dust lanes.
 
 The observed MIR excess at $> 5\, \mu$m is common in early-type galaxies, and is possibly due to the dust emission from circumstellar materials around AGB stars \citep{Shim11,Ko12}, residual star formation, or AGN. We examine the WISE 12 and 24 $\mu m$ images and find that the MIR light distribution is extended and follows the optical light distribution. Furthermore, the excess can be explained with a relatively old model that incorporates the AGB dust emission. On the other hand, the expected MIR flux from AGN activity as measured from the X-ray is two orders of magnitude lower than the excess value. Therefore, we suggest that the MIR-excess is due to the dust-emission from AGB circumstellar materials.
 
 
\section{Distance}

 Early-type galaxies exhibit a tight correlation between $r_{\rm eff}$, the surface brightness within $r_{\rm eff}$, $\avsbeff$, and the central velocity dispersion $\sigma_{0}$, known as the fundamental plane (FP) relation. Since $\avsbeff$ and $\sigma_{0}$ are quantities that can be measured without knowing  distance, FP can be used to measure distances to early-type galaxies. The FP can be expressed as,
 \begin{equation}
 \mathrm{log}(r_{\rm eff}\, h_{70}^{-1}/\mathrm{kpc}) = a\, \mathrm{log}(\sigma_{0}/\mathrm{km\, s^{-1}}) + b \, \avsbeff + c,
 \end{equation}
 where the coefficients $a, b$, and $c$ vary with wavelength \citep{Labarbera10, Jun08}. 
 
 Once $r_{\rm eff}$ is determined, the angular distance $d_{A}$ is calculated as
 \begin{equation}
 d_{A}  = (r_{\rm eff}\, h_{70}^{-1}/\mathrm{kpc}) \times \frac{206265}{\theta_{\rm eff}/\mathrm{arcsec}}.
 \end{equation}
 
 We gathered four independent measurements of  $\mathrm{log}(\sigma_{0}/\mathrm{km\,s^{-1}})$ from literature: $2.312\pm0.095$\citep{Carter88},
$2.237\pm0.013$ \citep{Beuing02}, $2.292\pm0.042$\footnote{\citet{Wegner03} listed a value within a 0.595 h$^{-1}_{100}$ kpc radius aperture, and this value is converted to the value for an aperture with 1/8 of $r_{\rm eff}$ using Eq.(1) of \citet{Cappellari06}.}  \citep{Wegner03}, 
and $2.212\pm0.045$ \citep{Ogando08}. 
 We adopt a weighted mean value, $\langle \mathrm{log}(\sigma_{0}/\mathrm{km\,s^{-1}}) \rangle =2.241 \pm 0.012$ of the four independent measurements.  
 Using the $\avsbeff$ values from Table \ref{tab:sbfit} after applying the Galactic extinction correction, and the FP coefficients from various sources, we derive distances to NGC 4993 in 7 different bands. Table \ref{tab:dist} summarizes the result.
  In total, 17 different estimates are derived. Two of the FP relations are based on $g,r,i$,and $z$ \citep{Bernardi03, Labarbera10} and for these cases, the $B, V, R$, and $I$ values are converted to $g,r$, and $i$ values while keeping the structural parameters in the corresponding bands. 
  Uncertainties in the distances are dominated by the intrinsic dispersion in the FP relation which has errors of about 7 to 12 Mpc \citep[or 23 \% of the derived value; e.g.,][]{Bernardi03}. The errors from the observed quantities amount to only $\lesssim$ 10 \% of the FP distance, and we ignore them. The uncertainty in the Hubble constant is only a few \% or less according to recent estimates \citep[e.g.,][]{Riess16},
   and can be neglected too. 
   We note that the rms dispersion of the 17 estimates is 5.3 Mpc, smaller than the FP distance error of each estimate. This suggests that the 17 estimates are not independent quantities (e.g., they share $\sigma_{0}$ value, and are based on an identical object), and the rms dispersion of 5.3 Mpc is an uncertainty related to the wavelength and the adopted FP relation parameters for each measurement. Therefore, we consider a typical intrinsic scatter in the FP relation along $\mathrm{log}(r_{\mathrm{eff}})$ of 0.09 dex (23 \% of the value) as our distance error.
    Considering these factors, we adopt 37.7 $\pm$ 8.7 $h^{-1}_{70}$ Mpc as the FP-based distance. Note that this is an angular diameter distance, and at $z=0.009783$ of NGC 4993 \citep{Levan17}, the luminosity distance is 2 \% larger at 38.4 $\pm$ 8.9 $h^{-1}_{70}$ Mpc.
  We also note that the mean angular diameter distance of 37.7 Mpc is accurate to about $5.3~ \rm{Mpc}/\sqrt{17} = 1.3$ Mpc in regard to 
 uncertainties due to wavelengths and FP parameter sets.

\begin{deluxetable}{c|cc}[h!]
\tablecaption{FP distance to NGC 4993\label{tab:dist}}
\tabletypesize{\scriptsize}
\tablecolumns{3}
\tablenum{3}
\tablehead{
\colhead{Filter} & \colhead{Distance} & \colhead{Mean distance}\\
\colhead{} & \colhead{(Mpc)} & \colhead{(Mpc)} \\
}
\startdata 
$B$ & 31.7\tablenotemark{a}, 34.7\tablenotemark{b}, 36.4\tablenotemark{c}, 41.7\tablenotemark{d} & 36.1 (4.2)\\
$V$ & 31.8\tablenotemark{a}, 33.0\tablenotemark{e}, 32.3\tablenotemark{f}, 41.8\tablenotemark{d} & 34.7 (4.7) \\
$R$ &  36.2\tablenotemark{a}, 41.3\tablenotemark{d}                                                                      & 38.8 (3.6) \\
$I$  & 36.9\tablenotemark{a},  44.6\tablenotemark{g}, 46.9\tablenotemark{d}                                   & 42.8 (5.2) \\
$J$ & 38.8\tablenotemark{a}                                          								   & 38.8        \\
$H$ & 39.6\tablenotemark{a} 													   & 39.6        \\
$K$ & 44.1\tablenotemark{a}, 28.4\tablenotemark{h}									   & 36.3 (11.1) \\
Average & 37.7 (5.3)                                                                                                                       &  38.2 (2.7) \\
\enddata
\tablenotetext{a}{\citet{Labarbera10}}
\tablenotetext{b}{\citet{Bender98}}
\tablenotetext{c}{\citet{delaRosa01}}
\tablenotetext{d}{\citet{Bernardi03}}
\tablenotetext{e}{\citet{DOnofrio08}}
\tablenotetext{f}{\citet{Jun08}}
\tablenotetext{g}{\citet{Scodeggio97}}
\tablenotetext{h}{\citet{Pahre98}}
\tablecomments{The numbers in the parentheses are the rms scatter of the corresponding quantities, not errors of the values. The FP parameters are adjusted so that $\sigma_{0}$ is the value inside an aperture radius of $\frac{1}{8}\,r_{\rm eff}$.}
\end{deluxetable}


\section{Discussion}

   Structural parameters suggest that NGC 4993 is an ordinary elliptical galaxy, having a SB profile consistent with that of a bulge-dominated galaxy with $n \sim 4$ and sitting right in the middle of the size-mass relation of local early-type galaxies \citep{Yoon17}. 
 The disk-like features and the dust lanes, common in post-merger ellipticals \citep[e.g.,][]{Kim13,Shabala12} suggest that NGC 4993 has gone through merging activities before. Indeed, this galaxy has been noted in the past as a ``shell elliptical'' \citep{Carter88}.

 One can derive the dynamical mass ($M_{dyn}$) of NGC 4993 and see how it compares with the mean stellar mass. Using the relation of $M_{dyn} = 5\,r_{\rm eff} \sigma_{\rm eff}^{2} /{\rm G}$ \citep{Cappellari06} where $\sigma_{\rm eff}$ is the $\sigma$ within an aperture with $r_{\rm eff}$ and rescaling $\sigma_{0}$ to $\sigma_{\rm eff}$ with Eq. (1) of \citet{Cappellari06}, we get $M_{dyn} = (5.2 - 7.8) \times 10^{10} \msun$ for $r_{\rm eff} = 3$ kpc. The value becomes smaller if we adopt $r_{\rm eff} = 2$ kpc in NIR. These values are in good agreement with the mean stellar mass.
  The properties of NGC 4993 
 are consistent with some of short GRB host galaxies, although our age estimate of $\gtrsim$3 Gyr lies at the old end. The trace of a minor merging event suggest that the BNS  system might have come from the merged galaxy. Another possible explanation for the BNS from a galaxy with very old stellar population is the dynamical origin \citep{Bae14} where the NS binaries can be formed within globular clusters via three body processes and eventually undergo merger after they get kicked out of the cluster.
  
  We find that the location of GW170817 is only about $10\farcs23 \pm 0\farcs08$ ($\sim$2 kpc) away from the center of NGC 4993, or 2/3 to 1 of $r_{\rm eff}$ in projected distance. Short GRBs tend to occur at outer regions of host galaxies (a median offset of 1.5 $r_{\rm eff}$; Fong \& Berger 2013). So far, about 25\% of short GRBs have been found at a projected offset of $< r_{\rm eff}$ from the host galaxy center \citep{Troja08,Fong13,Li16}, so the occurrence of the event near the center is not very unusual. 
   
   The luminosity distance of $38.4 \pm 8.9$ Mpc agrees with the distance estimate from the GW signal \citep[40$^{+8}_{-14}$ Mpc;][]{Abbott17b} as well as the previous estimate to the group distance \citep[$\sim40$ Mpc;][]{Sakai00}.  This independent assessment of the distance can possibly improve constraints on the GW source property such as the inclination of the binaries and eventually masses and spins.
   

\acknowledgments
This work was supported by the National Research Foundation (NRF) grants, No. 2017R1A3A3001362 and No. 2016R1D1A1B03934815, funded by the Korea government (MSIT). HML was supported by the NRF grant No. NRF-2016R1D1A1A02937544. CUL and SLK were supported by the KASI grant 2017-1-830-03. This research has made use of the KMTNet system operated by KASI, and the data were obtained at three host sites of CTIO, SAAO, and SSO. We thank the staffs of iTelescope.Net and the KMTNet sites for their excellent support.

\end{document}